Securing Big Data systems, A cybersecurity management discussion

Shakir A. Almasaari

Illinois Institute of Technology

Abstract

This paper explores the essential areas of cybersecurity management for big data systems. Big data platform stems its complexity from being a collection of interrelated non-standardized systems that interact with each other to process large data-sets. This complexity increases the chances of overlooking critical cybersecurity management practices. The paper discusses various security management for each part of big data systems. This includes security measures, standards, and methodologies. The primary objective is to highlight the essential segments of security management that are expected to be implemented flawlessly in big data systems.

Securing Big Data systems, A cybersecurity management discussion

In today's technology world, data are being generated in large volumes, with rapid velocity, and with a variety of format. These 3 Vs (volume, velocity, variety) are the main characteristics that describe big data. *Big Data* is a term, not a technology, refers to the extremely large data sets that require special handling. The National Institute of Standards and Technology (NIST) defines big data as "Big Data consists of extensive datasets primarily in the characteristics of volume, variety, velocity, and/or variability that require a scalable architecture for efficient storage, manipulation, and analysis" (2018, p. 6). Furthermore, researchers concluded that most of the data in the world has been generated in the last couple of years:

> Every day, we create 2.5 quintillion bytes of data. To put that into perspective, 90 percent of the data in the world today has been created in the last two years alone and with new devices, sensors and technologies emerging, the data growth rate will likely accelerate even more (IBM Marketing Cloud, 2017, p. 3).

*Big Data Systems* are defined as all the planning and management practices in addition to the technical components and infrastructure that are required to make use of the big data sets. Typically, big data systems stand on three interrelated layers: data storage and filesystems, data streaming engines, and computing nodes. The complexity and fragmentation of such interconnected systems raises the challenges of cybersecurity management. This paper sheds light on different areas of essential cybersecurity policies, standards, and best practices for big data environment. The first part will describe data lakes and their security management as well as discussing storage systems and distributed filesystems security management. The second part will discuss essential security management for data streaming engines. The last part will go through cybersecurity approaches that are critical for big data computing engines.

**Data Lake security management**

Up until recently, organizations were adopting Data Warehouses technologies for storing the organization's data assets. This approach helps data analysts to extract valuable insights that could support business decision processes. Data Warehousing architecture is suitable for capturing structured and filtered data that comes from transactional data sources such as Sales Transactions, Enterprise Resource Planning (ERP) and Customer Relationship Management (CRM) databases. Normally, data engineers would use Extract, Transform, and Load (ETL) tools to integrate the data from multiple data sources to the data warehouse databases after reshaping and cleaning the data. However, with the advent of big data, a different form of data is being generated such as: real time data-streams, web connected-devices, Internet of Things sensors, and social media feeds. These types of data have complex and dissimilar characteristics, thus, requiring different data management techniques and tools other than data warehousing. Moreover, most of the data are unstructured, has no clear schema, and require different integration approach to handle them. Consequently, enterprises started to look for a different data architecture that can fulfil such complex demands. A new data management approach known as *Data Lake* emerged to solve the problem. Data lakes is a data management architecture: a scalable vast pool of data repository that accepts data in its raw format from different sources. According to a research:

> A "Data Lake" is a methodology enabled by a massive data repository based on low cost technologies that improves the capture, refinement, archival, and exploration of raw data within an enterprise. A data lake contains the mess of raw unstructured or multi-structured data that for the most part has unrecognized value for the firm (Fang, 2015, p. 820).

The following figure shows the basic architecture of a data lake:

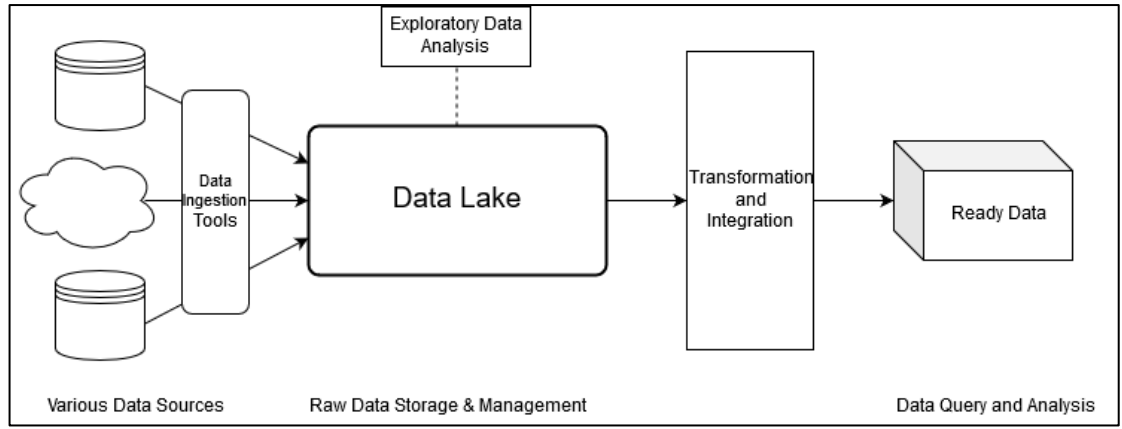

Figure 1

Data lake architecture provides many advantages. One advantage is that data lakes have no limits of storing data regardless of their characteristics, types or formats. Another advantage is that we don't need to process the data before storing it in a data lake. Storing raw data saves a lot of time and effort especially with data that have no clear schema. Thus, data cleaning jobs would be performed at the time when we only need to analyze the data. Moreover, data lake architecture is very flexible. It allows different tools with different technologies to access and query the data and gain insights. Finally, data lake architecture provides a unified point to view all the data across the enterprise, which achieves the concepts of breaking data-silos and data democratization (LaPlante & Sharma, 2016, p.3).

Although that data lake architecture is very empowering, it comes with risky tradeoffs. From data security point of view, having such unified large data repository requires effective security and access control practices. In general, the objective of data lake security is to guarantee guarding data by granting access to data lake components for only those who are authorized to perform specific job or duty.

**Developing a comprehensive data governance strategy**

Before implementing a data lake project, there has to be a holistic understanding of the multiple aspects of managing it. This includes developing strategic data governance policies, which without it, the data lake would lose the structure and becomes unmanageable (Kachaoui & Belangoour, p. 3). One data governance technique is to maintain the data lineage. Data lineage is a data lifecycle that helps in recognizing the origins of the data, who have accessed it, how it was changed, and where it transferred throughout its lifecycle. Data lineage is an important practice in data lakes due to the variety of sources that pours data into the lake. The lack of proper data governance approach for maintaining data lineage could make the data susceptible to malicious practices which leads to serious security issues. Thus, a good data governance approach ensures a comprehensive view of data for multiple parties without compromising data integrity, security, and reliability.

**Accessibility and users' privileges**

Data lakes are implemented using a wide array of software and hardware such as storage mediums and filesystem, operating systems, and computing servers. Therefore, there would be an imperative need to control access to the environment. In this section, the discussion will be focused on data lakes accessibility as a platform in general. The subsequent sections will cover the storage and filesystem security management practices in more depth.

Data lake architecture assumes that data would be frequently accessed from multiple users, programs, and Application Programming Interface (APIs) calls, internally and externally, with each access having its own different characteristics. Generally, users should only access the data to which they are granted permission at various granularity.

To organize the access to the platform, the management should implement Role Based Access Control (RBAC) policies. RBAC is an access-control type that aims to restrict persons' or software's access to specific resources, based on their roles in the organization. Applying RBAC is essential to ensure data lake security and dynamism. In a typical implementation, each business role will have its own RBAC policies. For examples, marketing business users' access-control policy should not overlap and should be separated from finance business users' ones and so on. Furthermore, in most cases, only C-level employees, with proper privileges, should have access to sensitive reports that are being generated in the data lake. From a different perspective, business users who use business intelligence tools to generate reports usually need to work with real time or near real-time data. Because, low data-latency is critical for them to generate accurate reports, their access-control policies should not hinder or impact the response time of the business intelligence servers. Other users, such as data scientists, usually need more access flexibility to explore and investigate the data to perform their duties effectively (LaPalnte & Sharma, 2016, p. 47).

**Storage systems security management**

Data storage layer is the last frontier to be defended. On this level, data security techniques are expected to be rigorously implemented to prevent intentional or unintentional data destruction. Generally, data in storage systems can be categorized in two types: data-in motion, and data at-rest. Data at rest is a term refers to the data that are inactive and do not transfer from a location to another regularly. Data at-rest are more vulnerable than data in-motion (data that transfer continuously) for couple of reasons: they provide a wealth of current and past data and they are vulnerable to be seized at huge volumes in a single attack (IEEE, 2014, p. 110).

Therefore, building a reliable storage system is challenging, especially in a big data environment, where there is always a growing amount of data and continuing demands for scalability. There are many safeguarding techniques known to security professionals such as storage traffic profiling, monitoring & reporting, and intrusion detection. However, this section will shed light on the essential data storage security management in terms of: data encryption, retention, and disposal.

**Storage media encryption**

Data cryptographic standards inevitably provide comprehensive methodology for organizations to protect its data assets and comply with multiple regulations requirements. It is worth mentioning that some regulatory bodies do not mandate data encryption for compliance, but they highly recommend it. For instance, the EU's General Data Protection Regulation (GDPR), in their data storage security compliance guide, mentions that data encryption is not compulsory for satisfying the EU GDPR requirements, but it does help organizations in complying with the GDPR data security overall guidelines (Rubens, 2019).

One storage security standard that covers data encryption is the ISO/IEC 27040. It provides detailed guidelines on securing data storage and encrypting the data whether it is at-rest or in-motion. The ISO/ICE 27040 standards emphasize on the effectiveness of *near data-origin* encryption to enforce strong control on the data. According to SNIA report:

> While ISO/IEC 27040 recognized that data is best encrypted as close as possible to the origin of that data, it also recognizes that encryption used near the point of storage provides an effective mechanism to combat situations where control of the media is lost (e.g., storage media recycled, discarded, etc.). In these cases, technologies such as self-

    encrypting drives (SED), controller-based encryption, and tape encryption may be

    especially useful (Storage Networking Industry Association, 2015, p. 14).

Despite the type of encryption algorithm, there are general recommendations that the enterprise should consider. First, as per the ISO/IEC 27040 standards, the minimum recommended encryption strength is 128 bits. Second, multi-layer encryption can be implemented on other higher layers, such as the application and computing layers. Third, all encryption processing (such as key generation and activation) should be properly logged to enable security auditing on the encryption activities. Finally, encryption should not be the solely practice in protecting storage systems.

  Other storage-level protection approaches work collectively with data encryption in protecting the storage system (Storage Networking Industry Association, 2015, p. 14). Organizations developing big data systems could consider the IEEE1619 standards for storage layer encryptions. These standards provide encryption methods and algorithm recommendations for securing data on a storage media. The IEEE1619 standards achieve the interoperability of data encryption while it is being transmitted between storage media. According to Hughes & Cole "Using this standard, companies can provide interoperable encryption hardware and software to protect information while it is being sent over Fiber Channel, SCSI, or any other means to the storage device." (2003, p.124).

  Because big data systems utilize the distributed computing architecture, data will be in continues movement between different storage and computing nodes. So, the IEEE P1619 provides suitable encryption standards for distributed systems. For example, implementing the IEEE P1619 standards protects the back-end storage systems while data moves back and forth against different types of threats such as sniffing and man-in-the-middle attacks (Hughes & Cole, 2003, p. 125).

**Data retention policies**

Proper data retention techniques are critical especially for document-based information. This kind of information represents documented evidence of business transactions, regulatory compliance, and various aspects that touches the organization's liability and reasonability (Storage Networking Industry Association, 2018, p. 20). Therefore, data retention policies should be determined based on business needs and legal requirements and should cover many important elements.

One element is determining the period of how long the organization should keep the information and what are the archiving procedures to ensure data confidentiality. Another element is that the data retention policy should distinguish between various retention periods. Evidently, not all the data are needed to be retained permanently. The retention periods could be categorized to: short, medium, and long periods. Short to medium data retention periods are usually developed to satisfy legal and regulatory requirements to avoid problematic liabilities on the business. However, long data retention periods tend to expose the data to attackers. Because data archives reside in archival systems for a very long time, they become susceptible to cyber-attacks for a longer time (Storage Networking Industry Association, 2018, p. 15).

That said, the ISO/IEC 27040 standards provide data retention recommendations that help mitigate the risks accompanied with long data retention periods. The following is a summary of the ISO/IEC 27040 standards recommendations:

- Data integrity in the archival systems should not be neglected and should be frequently examined.

- Migrating data archives to newer storage systems should not suppress any previously used effective security capabilities.

- Because data in archives tend to last for very long time, it might out-live the data governance roles including data-stewardship roles. Therefore, storage systems should be able to enforce user authentication for new users and assigning the proper resources accessibility.

- Security techniques, such as data encryption and decryption, should always be anonymous and hidden to the user who creates the data.

- To detect slow attacks and to better avoid future attacks, security functions and processes should always be logged, tracked, and maintained for a long time.

- The system should be able to deal with data breaches instantly with the capability of keeping records of previous breaches to help in taking proper actions in case if a new incident occurs.

- Storage enhancement mechanisms (such as data compression) should neither compromise security capabilities nor data integrity (Storage Networking Industry Association, 2018, pp. 15–16).

**Data sanitization and disks disposal**

In storage systems, disks are expected to be replaced more frequently than any other hardware part. Hardware lifespan, generally, is three to five years; after that, they become obsolete. The usual routine in data centers is to remove these obsolete hardware parts and git rid of them systematically. Because storage mediums contain data, effective disks sanitization is imperative before storage media disposal in order to maintain data confidentiality.

To show how some businesses neglect proper disk sanitization, a study conducted by Blancco Technology Group's specialists in data erasure, found that 78 presents of 200 secondhanded hard disks drives bought from eBay and Craigslist were not properly formatted and sanitized. The company, indeed, were able to recover those disks and found that 67 percent of the disks were containing personal data such as images, locations, and financial and 11 percent contained company data (Kan, 2016). Such data leak is disastrous, and the negative impact could be exponential in big data environment due to the huge number of disks used in storage systems. Therefore, organizations should have a clear policy of data disposal that satisfies business needs and legal compliance as well as protects data confidentiality. The following points are essential for disks disposal policy:

- There should be a designated department or team responsible for the disk disposal process.
- The disposal team should adhere to the industry standards in disks sanitizations or destruction to insure best practices applied.
- A sanitization software should be used to remove all data from disks' sectors by applying zero-filled blocks overwriting mechanism that meets the Department of Defense standards.
- If feasible, hard drives can be demolished using proper mechanisms such as disks crushing and destruction (SysAdmin, Audit, Network and Security (SANS) Institute, 2014, p. 2).

**Distributed filesystems security management**

Distributed Filesystems are filesystems designed to manage and utilize the resources in a distributed computing environment. This includes enabling multiple nodes to communicate and share data via shared network. Distributed filesystems are also designed to cope with distributed computing needs such as: scalability, fault-tolerance, and high availability. In terms of security management, distributed filesystems in a typical deployment, use networks protocols and access lists to restrict access to system resources. Access to the distribute filesystem could include: listing available directories, mounting and unmounting devices and folders, reading and writing files, running executable files, and many more. Using security protocols can help protect distributed filesystems by enforcing client-server shared authentication as well as clients-authorization to permit filesystem access (Zarei, Asadi, Nourizadeh, & Begdillo, 2008, p. 305). Essentially, there are two important parts that distributed filesystems should implement: One is using authentication layer; and two is using an authorization mechanism.

First, ensuring that the communication between the nodes in the cluster is authenticated using secured-communication protocol such as Kerberos. Kerberos is a network authentication protocol that ensures a secure and authenticated communication between multiple-nodes in a client-server environment using secret-key cryptography mechanism. It was designed to allow communicating through insecure network (e.g. the internet) so every node in the network can verify its identity to the others securely (Massachusetts Institute of Technology, 2019).

Second, using users-level authorization mechanism. This can be implemented via enforcing Access Control Lists (ACLs) policies and Portable Operating Interface (POSIX for UNIX like systems) style permissions on filesystem's directories and files. This practice can protect the filesystem from unauthorized access.

**Data Streams security management**

*Data Streams* is a continuous flow of data from a point (data generators) to another point (data receiver or consumer) in a non-stop fashion. *Data Streaming* is a mechanism for managing and processing different types of data streams from multiple sources. Baird defines data streaming as:

> Data streaming is a strategy employed when the source of information generates data on a continuous basis and near real-time updates are required to allow analysts a more recent view of data (usually aggregated) by which they make decisions (Baird, 2019).

Data streaming has become a central technology in big data and the Internet of Things (IoT) projects because it provides real-time processing framework for data streams. Some use cases that heavily rely on real-time data processing, such as stock market monitoring, fraud detection, tracking devices, social media feeds analysis, cybersecurity intelligence, patients tracking, and multimedia services. From a business perspective, streaming analysis has a huge potential. According to PR-Newswire, streaming analytics market was $5021.1 Million in 2015 and this number is expected to grow up to around $2 billion by 2020 (PR Newswire , 2015). This emphasizes the need to build secure data streaming systems.

Data streams are managed and controlled via a framework known as Data Stream Management Systems (DSMS) and it is typically implemented in a distributed computing platform. Software engineers use specific APIs and programming libraries to write custom applications to initiate, analyze, and manage data streams. From a cybersecurity point of view, the complexity in protecting big data streams resides in ensuring that data are not being intercepted or corrupted while it pours in rapidly and continuously from different sources. Therefore, access lists policies and data streams encryption mechanisms must be present.

First, access lists on the Data Stream Management Systems (DSMS) provide authorization layer to filter out data-fetching processes from or to the streaming platform. There are three role-based access control models that can be used to enable access control on data streams: preprocessing, postprocessing, and query rewriting. Preprocessing enforces access control to prevent the query from fetching data from unauthorized tuples in advance. On the contrary, postprocessing filters out the unauthorized tuples from the query result after executing the user query and before submitting the results to the data stream. Query rewriting, however, takes a different approach. When a user submits a query, an optimization tool, called query rewriter, verifies whether the query can be executed fully or partly based on the authorization list. If the query has partial authorization, it would be rewritten by the tool to comply with the access list policy and then would be executed. This mechanism ensures the output of the authorized data regardless of the nature of the submitted query (Carminati et al., 2010, p. 22).

Second, data streams can be encrypted from different levels to ensure end to end security. There are mainly two levels of data streams encryption: communication level encryption, and API (program code) level encryption. For encrypting the communication in a data stream, an organization could use the Hyper Text Transfer Protocol Secure (HTTPS) over Secure Socket Layer (SSL) or over Transport Layer Security (TLS) during data transmission. This is a well adopted approach among security institutes. According to the Federal Chief Information Officers council (CIOs): "The internet's standards bodies, web browsers, major tech companies, and the internet community of practice have all come to understand that HTTPS should be the baseline for all web traffic." (Federal Chief Information Officers (CIOs), n.d.).

Along with encrypting the communication layer, organizations could use data encryption on the code level if necessary. When software engineers develop a data streaming application, they could utilize specific libraries and APIs to enforce encryption on the data that are being generated by the application. This encryption approach depends on the business requirements. In some cases, data streams are processed within the organization's local network and are not exposed to any external resources. In this model, streams encryption might not be effective. Therefore, the organization should conduct proper assessment regarding whether that data streams will be exposed to external environment or not. Encryption process should not be enforced arbitrarily because it comes with a technical overhead and performance impacts especially for real-time data processing.

Data streams might look challenging to secure, but it also could help implement sophisticated security features. For instance, data-security analytics framework can detect security threats in real-time and take proper actions automatically, efficiently, and quickly. This is a novel threat detection approach that is being implemented widely by utilizing big data streaming engines. Basically, a streaming analytics engine examines the network traffic in real-time or near real-time, looking for anomalies and suspicious behaviors that could indicate an attack. Then after, the framework would take actions that are programmed to do, such as: isolating the threat, locking down accounts access, verifying the authentication again, blocking the communication, and notifying security personnel ("Security Concerns in Streaming Analytics from Data Processing Centers," 2017). This approach enables organizations to gain a comprehensive knowledge about the data traffic and to enforce high level of cybersecurity practices.

### Big data computing engines security management

Computing engines are central to big data framework where all the massive data processing and analytic tasks occur. Typically, computing engines generate meaningful information to applications and users in various forms, such as: dashboards, reports, and files. Thus, computing engines are vulnerable to cyberattacks since they provide rich and valuable information for the attackers. The following discussion will cover the essential security management practices for big data computing layer with providing examples on Apache Hadoop security.

Hadoop is a widely used open-source big data framework for running applications on a cluster of commodity hardware and processing data on scale. In its core, Hadoop consists of Hadoop Distributed Filesystem (HDFS), MapReduce – A parallel programming paradigm for distributed computing, and other pluggable related components, mostly are other Apache open source projects, that provide higher level of analytics and data processing abstraction such as: Apache Hive, and Apache Pig. Figure 2 shows the overall Hadoop architecture:

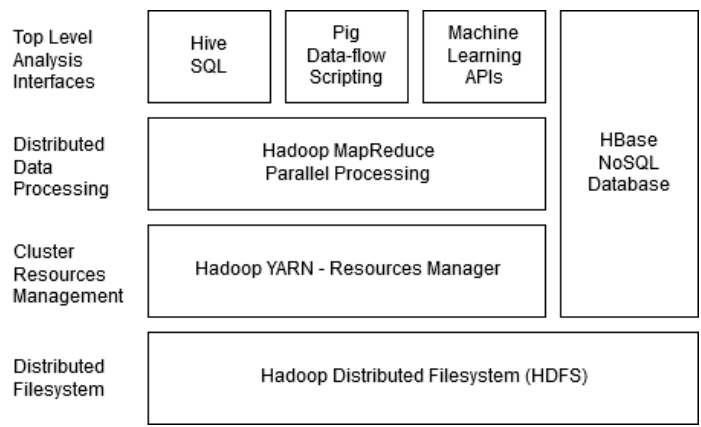

Figure 2

From the figure we can notice that the platform is composed of different components that interact with each other. One of the challenges in securing such processing framework is the absence of security standards that govern all these different components. It is also important to know that most of the available security solutions for Hadoop and its other components are fragmented; thus, securing such a platform require proper security planning (Gaddam, 2015, p. 2). The planning should cover at least the essential management of identity & access management, network security, and data protection.

First, there should be an identity & access management mechanism enforced on all the computing cluster components. Accessibility policies should be the central point to manage users access to data. The following are critical points that should be considered when developing identity & access management policies for big data computing nodes:

- Access policy should be developed based on data-access not on the access scheme.
- Access to the computing cluster should be based on Role-Based Access Control (RBAC) which delivers fine-tuned authorization level based on roles not users.
- Attribute based access control and data tags help ensure proper access permissions throughout all the phases of the data lifecycle.
- Limiting data access using data metering techniques for a user or application when reaching predefined threshold ensures the dynamism of the cluster and prevent data clogging (Gaddam, 2015, p. 5).

Second, securing computing cluster's network is an important aspect to prevent data disclosure. In Hadoop, a cluster consists of multiple nodes with a Master-Workers relationship (NameNode and DataNodes) connected via shared network. Data in Hadoop cluster are always in transit between the different nodes. Therefore, it is important to enable proper network security practices such as:

- Enabling secured communication protocols (such as HTTPS over TLS) to enable packet level encryption within all the cluster's nodes.
- Using LDAP over SSL (LDAPS) communication protocol to eliminate sniffing attacks.
- System administrators should encrypt data-transmission between all the cluster's components, such as resource management and processing servers and data sources (Gaddam, 2015, p.6).

Finally, application level cryptography and data masking help protect data while it is being processed in the computing layer. Application level cryptography refers to the ability to encrypt data in specific data-fields. It provides high-level granularity, but it requires manual processing to determine what data-field to be encrypted (Gaddam, 2015 p.5). Moreover, data masking is a technique for obscuring the data based on specific configurations to prevent unauthorized access to sensitive data in the cluster. Data masking does not change the original data, they only get obfuscated when presented to the user. Developers and data analysts in Hadoop environment require frequent access to data to perform their duties. Sometimes, it is not secure to expose real data to any user. In this case, data masking mechanism guarantees data protection without interrupting users' duties.

## Conclusion

To sum up, big data systems require well-planned cybersecurity management due to their heterogeneity, complexity, and fragmentation. The case of deploying a data lake architecture to serve as a big data platform requires comprehensive data governance strategy and users-accessibility policies to maintain the structure of the data lake. For storage systems security, cryptography methods provide comprehensive data protection and helps to comply with various regulations. Moreover, data retention practices are critical in big data systems because of the large data volumes that require proper archiving. Data retention periods policies are also important for various business responsibilities and liabilities. Storage disposal practices should follow strict process to ensure data confidentiality. Besides, big data streaming framework should implement access list policies and streams-encryption to guarantee end-to-end security. As was previously stated, in big data computing engines data masking and identity & access management should be implemented across all the cluster's nodes to prevent sensitive data exposure.

SECURING BIG DATA SYSTEMS                                                                21IEEE. (2014). *IEEE Standard for Encrypted Storage*. Retrieved from

    https://ieeexplore.ieee.org/stamp/stamp.jsp?arnumber=1362602

Kachaoui, J., & Belangour, A. (2019). Challenges and Benefits of Deploying Big Data Storage

    Solution. *Proceedings of the New Challenges in Data Sciences: Acts of the Second*

    *Conference of the Moroccan Classification Society Article No. 22- SMC '19*.

    https://doi.org/10.1145/3314074.3314097

Kan, M. (2016, June 28). Used hard drives on eBay, Craigslist are often still ripe with leftover

    data. Retrieved October 16, 2019, from https://www.pcworld.com/article/3089343/resold-

    hard-drives-on-ebay-craigslist-are-often-still-ripe-with-leftover-data.html

LaPlante, A., & Sharma, B. (2016). *Architecting Data Lakes* (First Edition). Sebastopol, CA,

    United States of America: O'Reilly Media, Inc.

Massachusetts Institute of Technology. (2019, January 9). Kerberos: The Network Authentication

    Protocol. Retrieved October 26, 2019, from https://web.mit.edu/kerberos

National Institute of Standards and Technology. (2018). *NIST Big Data Interoperability*

    *Framework: Volume 1, Definitions*. Retrieved from

    https://bigdatawg.nist.gov/_uploadfiles/NIST.SP.1500-1r1.pdf

PR Newswire . (2015, June 22). Global Streaming Analytics Market by Verticals - Market

    Forecast & Analysis 2015 - 2020. Retrieved October 15, 2019, from

    https://www.prnewswire.com/news-releases/global-streaming-analytics-market-by-

    verticals---market-forecast--analysis-2015---2020-300102535.html

SECURING BIG DATA SYSTEMS                                                                      22Rubens, P. (2019, May 10). How to Comply with GDPR. Retrieved October 15, 2019, from

    https://www.esecurityplanet.com/network-security/how-to-comply-with-gdpr.html

Security Concerns in Streaming Analytics from Data Processing Centers. (2017, August 25).

    Retrieved September 17, 2019, from https://imaginenext.ingrammicro.com/data-

    center/security-concerns-in-streaming-analytics-from-data-processing-centers

Storage Networking Industry Association. (2015). *Storage Security: Encryption and Key*

    *Management*. Retrieved from

    https://www.snia.org/sites/default/files/technical_work/SecurityTWG/SNIA-Encryption-

    KM-TechWhitepaper.R1.pdf

Storage Networking Industry Association. (2018). *Storage Security: Data Protection*. Retrieved

    from https://www.snia.org/sites/default/files/security/SNIA-Data-Protection-

    TechWhitepaper.pdf

SysAdmin, Audit, Network and Security (SANS) Institute. (2014). *Technology Equipment*

    *Disposal Policy*. Retrieved from https://www.sans.org/security-resources/policies/server-

    security/pdf/technology-equipment-disposal-policy

Zarei, B., Asadi, M., Nourizadeh, S., & Begdillo, S. J. (2008). A Novel Security Schema for

    Distributed File Systems. *Advances in Computer and Information Sciences and*

    *Engineering*, 305–310. https://doi.org/10.1007/978-1-4020-8741-7_56